\documentclass[letter]{aa} % for the letters %
\usepackage{natbib}
\bibpunct{(}{)}{;}{a}{}{,}
\usepackage[pdftex]{graphicx}
\usepackage{txfonts}
\begin{document}
\title{Direct detection of a substellar companion\linebreak~to the young nearby star PZ\,Telescopii \thanks{Based on observations obtained at Paranal Observatory in ESO programs 079.C-0908(A), 083.C-0150(B), 085.C-0012(A)}}

\author{M. Mugrauer\inst{1}
\and
N. Vogt\inst{2}$^{,}$\inst{3}
\and
R. Neuh\"{a}user\inst{1}
\and
T.O.B. Schmidt\inst{1}}

\offprints{Markus Mugrauer \email{markus@astro.uni-jena.de}}

\institute{Astrophysikalisches Institut und Universit\"{a}ts-Sternwarte Jena, Schillerg\"{a}{\ss}chen 2, 07745 Jena, Germany
\and
Departamento de F\'isica y Astronom\'ia, Universidad de Valpara\'iso, Avenida Gran Breta\~na 1111, Valpara\'iso, Chile
\and
Instituto de Astronom\'ia, Universidad Catolica del Norte, Avda. Angamos 0610, Antofagasta, Chile \\}

\date{Received 2010 August 04; accepted 2010 August 25}

\abstract{}{We study the formation of substellar objects (exoplanets and brown dwarfs) as companions to young nearby stars.}{With high contrast AO imaging obtained with NACO at ESO's VLT we search for faint companion-candidates around our targets, whose companionship can be confirmed with astrometry.}{In the course of our imaging campaign we found a faint substellar companion of the nearby pre-main sequence star PZ\,Tel, a member of the $\beta$ Pic moving group. The companion is 5-6\,mag fainter than its host star in JHK and is located at a separation of only 0.3\,arcsec (or 15\,AU of projected separation) north-east of PZ\,Tel. Within three NACO observing epochs we could confirm common proper motion ($>39\sigma$) and detected orbital motion of PZ\,Tel\,B around its primary ($>37\sigma$). The photometry of the newly found companion is consistent with a brown dwarf with a mass of 24 to $40\,M_{Jup}$, at the distance (50\,pc) and age (8$-$20\,Myr) of PZ\,Tel. The effective temperature of the companion, derived from its photometry, ranges between 2500 and 2700\,K, which corresponds to a spectral type between M6 and M8. After $\beta$\,Pic\,b, PZ\,Tel\,B is the second closest substellar companion imaged directly around a young star.}{}


\maketitle

\section{Introduction}

Since several years we perform a high contrast imaging search for substellar companions of young nearby stars using the adaptive optics and infrared imager NACO, which is installed on the Nasmyth-platform of UT4 (Yepun) at ESO's Paranal Observatory, in Chile. We observed PZ\,Tel (HD\,174429, HIP\,92680), which is a solar analog pre-main sequence star of spectral type K0Vp, at a distance of $49.7\pm2.9$\,pc \citep{perryman1997}; PZ\,Tel has a mass of 1.02$\pm$0.04\,$M_{\odot}$ \citep[see][]{allende1999} and \cite{randich1993} first reported about the fast rotational velocity ($vsin(i)=70$\,km/s), and the Lithium abundance ($N(Li)=3.9$) of this star, both important youth indicators. These results were then confirmed later on by others \citep[see e.g.][]{soderblom1998, rocha2002, barnes2000}. Due to its space motion ($[U,V,W]=[-10.6,-15.4,-7.9]$\,km/s) together with its other youth indicators, \cite{zuckerman2001} and \cite{torres2006} proposed that PZ\,Tel is a member of the $\beta$\,Pic moving group, with an age of $12^{+8}_{-4}$\,Myr. \cite{tetzlaff2010} determined the age of PZ\,Tel by comparing its luminosity and temperature to several theoretical model isochrones yielding an age of $12.8\pm2.2$\,Myr, consistent with the general age estimate for the $\beta$\,Pic moving group. \cite{rebull2008} reported about the detection of excess emission in the spectral energy distribution of PZ\,Tel at 70\,$\mu$m with MIPS/Spitzer. These observations indicate that the star is surrounded by a cold ($\sim$\,40\,K) low-mass dust disk with a total mass of only $\sim$\,0.3\,$M_{Moon}$, which extends from its inner border at a separation of about 35\,AU up to $\sim$\,200\,AU at its outer edge.

\section{Observations: Astrometry and Photometry}

We have imaged PZ\,Tel the first time in September 2009 with NACO's S13 optics in the $\rm K_{s}$-band. The imaging-setup of all observations is listed in Tab.\ref{table_log}.

\begin{table}[htb] \caption{Observation-log for all imaging data presented in this letter, with the detector integration time (DIT), the number of detector integrations per image (NDIT), as well as the number of frames ($N$) taken in total.}
\begin{tabular}{l|c|l|c|c|c}
\hline\hline
optics           & epoch      & filter                        & $DIT$ [s] & $NDIT$ & $N$\\
\hline
S27\footnotemark & 2007/06/13 & $\rm K_{s} + ND_{Short}$      & 1.000     & 50     & 5\\
S13              & 2009/09/28 & $\rm K_{s}$                   & 0.347     & 122    & 42\\
S13              & 2010/05/05 & J                             & 0.347     & 100    & 10\\
S13              & 2010/05/05 & $\rm H + ND_{Short}$          & 5.000     & 11     & 10\\
S13              & 2010/05/05 & $\rm K_{s}+ND_{Short}$        & 5.000     & 11     & 10\\
S13              & 2010/05/06 & $\rm J + ND_{Short}$          & 5.000     & 11     & 10\\
S13              & 2010/05/07 & $\rm J + ND_{Short}$          & 5.000     & 11     & 30\\
S13              & 2010/05/07 & $\rm H + ND_{Short}$          & 5.000     & 11     & 21\\
S13              & 2010/05/07 & $\rm K_{s} + ND_{Short}$      & 5.000     & 11     & 9\\
\hline
\end{tabular}

\label{table_log}
\end{table}
\footnotetext{Public data from the ESO data archive, taken in ESO program 079.C-0908(A).}

\begin{figure} [htb]
\resizebox{\hsize}{!}{\includegraphics{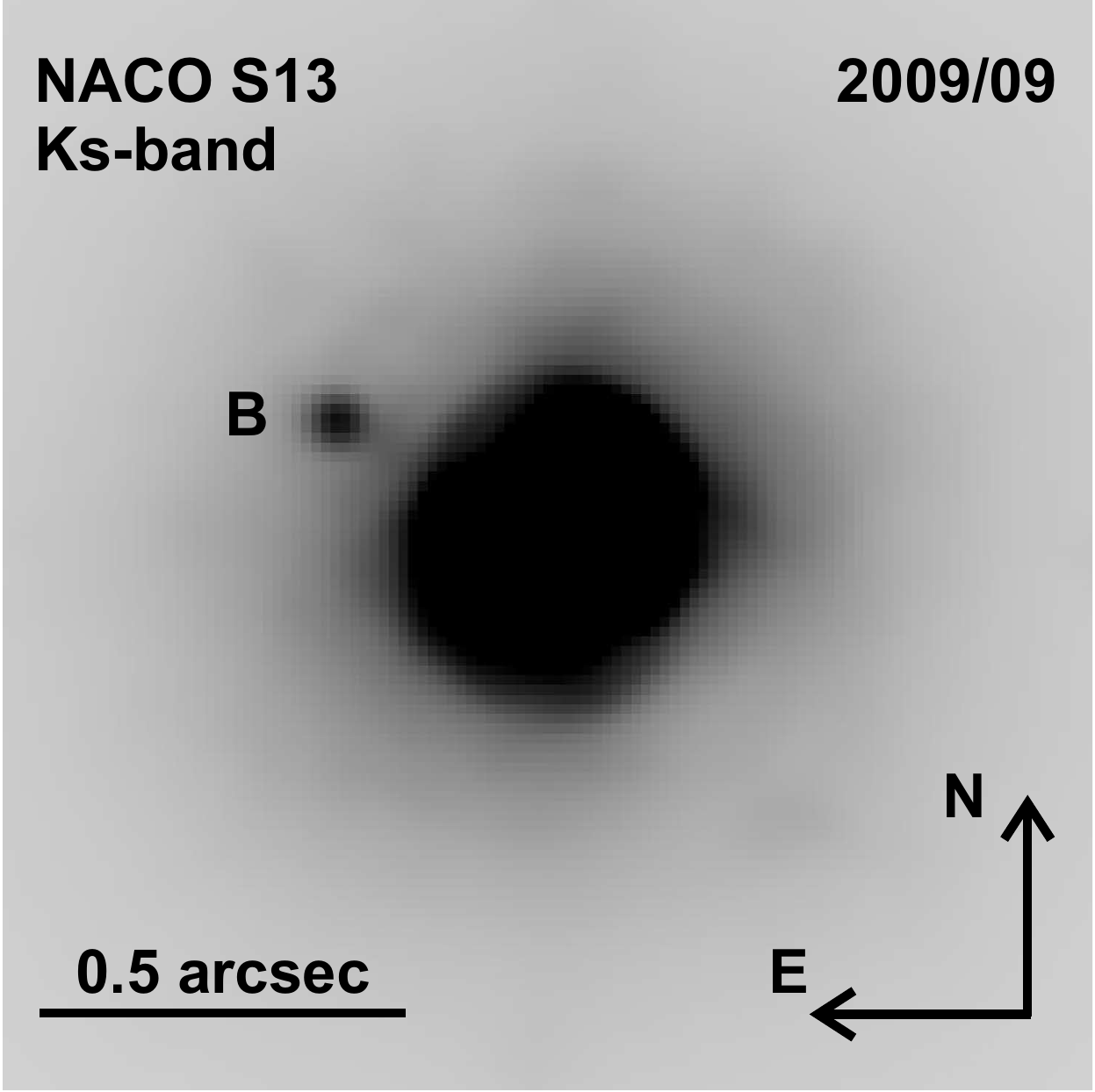}}\caption{This pattern shows our deep NACO image of PZ\,Tel together with its faint ($\Delta \rm{K_{s}}>5$\,mag) companion-candidate, taken in September 2009. The total integration time is 29.6\,min in the $\rm K_{s}$-band.}\label{pic}
\end{figure}

Due to the brightness of PZ\,Tel ($K_{s}\sim6.4$\,mag), we always used the shortest possible detector integration time ($DIT=0.347$\,s) to limit saturation only on the central few pixels of the point spread function (PSF) of the bright star, and 122 such integrations were then averaged to one image. In order to subtract the bright sky background in the $\rm K_{s}$-band, the telescope position was then moved between the individual images, and 42 images were taken in total, within a jitter-box with a width of 7\,arcsec. Internal lamp flats, as well as skyflats were used to calibrate the individual pixel sensitivity of NACO's infrared detector. The background estimation and subtraction, as well as the flat-fielding of all images, was then achieved with the ESO package \textsl{ECLIPSE}\footnote{ESO C Library for an Image Processing Software Environment} \citep{devillard2001}, which finally also combined all images, using a shift+add procedure. The fully reduced NACO image, taken in September 2009, is shown in Fig.\,\ref{pic}.

In our deep NACO observation beside the bright central star PZ\,Tel we detected a faint ($\Delta K_{s}>5$\,mag) companion-candidate, which is located only $\sim$\,0.3\,arcsec north-east of the star. In order to test whether this object is a real companion of the star or only a background source, which is just randomly located close to but far behind PZ\,Tel, further observing epochs are needed, in which the candidate is imaged together with PZ\,Tel. Therefore, we searched first in the ESO data archive for additional NACO observations of PZ\,Tel, and found a data-set in which the candidate is clearly detected. The public archival data were taken in June 2007 with NACO's S27 optics through the $\rm K_{s}$-band filter in combination with NACO's neutral density filter $\rm ND_{Short}$ (transmission of about 1.4\,\%).

To determine the accurate astrometry of the candidate relative to PZ\,Tel, images of astrometric calibrators have to be taken in each observing run. Therefore, in September 2009, we observed the globular cluster 47\,Tuc for which precise HST astrometry is available for several of its members. We determined the pixel scale $PS$ and the position angle $PA$ of NACO's S13 optics, which is listed in Tab.\,\ref{table_astrocal} together with the astrometric calibration of the archival data, taken from \cite{chauvin2010}.

\begin{table}[htb] \caption{The astrometrical calibration of NACO for all observing epochs whose data are presented in this letter. The pixel scale $PS$ and the detector position angle $PA$ are listed with their uncertainties.}
\begin{tabular}{l|c|c|c}
\hline\hline
optics                    & epoch   & $PS$ [mas/pixel]   & $PA$ [$^{\circ}$]\\
\hline
S27\footnotemark          & 2007/06 & $27.01\pm0.05$     & $-0.06\pm0.15$\\
S13                       & 2009/09 & $13.234\pm0.018$   & $+0.42\pm0.10$\\
S13                       & 2010/05 & $13.231\pm0.020$   & $+0.67\pm0.13$\\
\hline
\end{tabular}
\label{table_astrocal}
\end{table}
\footnotetext{Astrometric calibration of public archival data from ESO program 079.C-0908(A), listed in \cite{chauvin2010}}

\begin{figure} [htb]
\resizebox{\hsize}{!}{\includegraphics{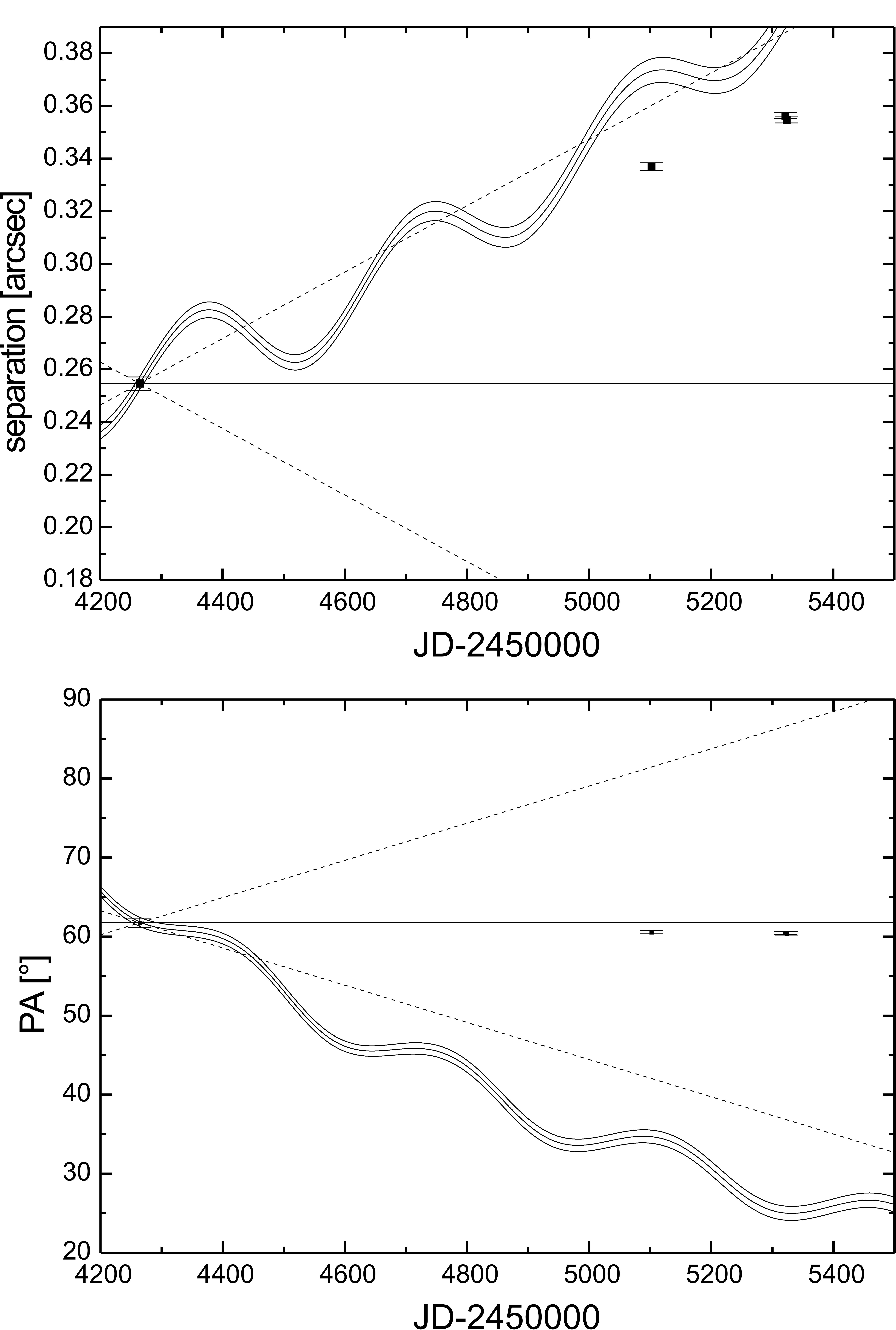}}\caption{The angular separation and position angle of the detected candidate for all observing epochs. The wobbled lines illustrate the expected change of both quantities in the case that the candidate is a non-moving background object. The dashed straight lines show the expected maximal change of both parameters for an object on a gravitationally bound orbit around PZ\,Tel.}\label{seppa}
\end{figure}

With the given astrometric calibration we could then determine the relative astrometry of the companion-candidate, i.e. its angular separation ($sep$) and position angle ($PA$) to PZ\,Tel, in all available observing epochs. Due to the small separation of the candidate from the much brighter star, the PSF of PZ\,Tel was always removed in all images with spatial filtering, to determine the relative astrometry of the candidate. All astrometric measurements are summarized in Tab.\,\ref{table_seppa} and are illustrated in Fig.\,\ref{seppa}.

\begin{table}[htb] \caption{The angular separations ($sep$) and position angles ($PA$) of the detected close companion-candidate relative to PZ\,Tel for all NACO observing epochs. In the columns $sep_{~bg}$ and $PA_{~bg}$ we show the expected separations and position angles in the case that the detected candidate would be a non-moving background object. The significance level on which this background hypothesis can be rejected is listed in the column $s_{bg}$, the one of detected orbital motion in the column $s_{orb}$.}
\begin{tabular}{c|c|c|c|c}
\hline\hline
optics \& epoch      & $sep$ [mas]       & $sep_{~bg}$ [mas] & $s_{bg}$ [$\sigma$] & $s_{orb}$ [$\sigma$]\\
\hline
S27 2007/06/13 & $254.6\pm2.5$     & ---               & --- & ---\\
S13 2009/09/28 & $336.9\pm1.5$     & $373.8\pm4.7$     & 8   &  28\\
S13 2010/05/05 & $356.3\pm1.1$     & $394.6\pm5.3$     & 7   &  37\\
S13 2010/05/07 & $354.8\pm1.3$     & $395.4\pm5.3$     & 7   &  36\\
\hline
optics \& epoch      & $PA$ [$^{\circ}$] & $PA_{~bg}$ [$^{\circ}$] & $s_{bg}$ [$\sigma$] & $s_{orb}$ [$\sigma$]\\
\hline
S27 2007/06/13 & $61.74\pm0.58$    & ---               & --- &---\\
S13 2009/09/28 & $60.55\pm0.22$    & $34.77\pm0.84$    & 30  &  2\\
S13 2010/05/05 & $60.47\pm0.21$    & $24.89\pm0.89$    & 39  &  2\\
S13 2010/05/07 & $60.41\pm0.21$    & $24.88\pm0.89$    & 39  &  2\\
\hline
\end{tabular}
\label{table_seppa}
\end{table}

The comparison of our NACO data from September 2009 with the archival data from June 2007 already proofs that the detected companion-candidate is clearly not a non-moving background source. The expected separations and position angles for such an object can be derived (see Tab.\,\ref{table_seppa}, and Fig.\,\ref{seppa}) from the relative astrometry of the candidate, the given epoch differences, as well as the proper and parallactic motion of PZ\,Tel, well known from Hipparcos measurements\footnote{$\mu_{\alpha}cos(\delta)=16.64\pm1.32$\,mas/yr, $\mu_{\delta}=-83.58\pm0.87$\,mas/yr, \linebreak and $\pi=20.14\pm1.18$\,mas \citep{perryman1997}}. We find that the position angle of the candidate only slightly decreases between both observing epochs by $1.19\pm0.62$\,$^{\circ}$ ($\dot{PA} = -0.5\pm0.3$\,$^{\circ}$/yr), while a decrease of about 27\,$^{\circ}$ is expected for a non-moving background source. Hence, we can reject the background hypothesis for the detected candidate already from its position angle measurements (on a significance level $> 39\,\sigma$). While the change in the position angle is relatively small, the separation of the candidate significantly increases between both observing epochs by $82.3\pm2.9$\,mas ($\dot{sep} = 35.8\pm1.3$\,mas/yr). However, even this high relative motion can be explained with orbital motion around PZ\,Tel (see section 4 for discussion). Therefore, we conclude that the detected candidate is a real companion of PZ\,Tel, and we will denote it as PZ\,Tel\,B, from now on.

In order to follow the orbital motion of PZ\,Tel\,B around its primary we observed the PZ\,Tel system in May 2010 again with NACO in the $\rm K_{s}$-band. For the astrometric calibration images of the globular cluster 47\,Tuc were taken, as well as in each night images of the wide binaries HIP\,42758, HIP\,45072, HIP\,46657, HIP\,50602. The instrument is found to be stable within the individual nights in May 2010. In addition, to our $\rm K_{s}$-band observations we also obtained images in the J-, and H-band to determine the infrared photometry of PZ\,Tel\,B. The magnitude difference between PZ\,Tel\,B and its primary is measured always after the subtraction of the PSF of the bright star in all NACO images, and is summarized in Tab.\,\ref{table_photo}.

\begin{table}[htb] \caption{Photometry of PZ\,Tel\,B. The used filters, as well as the measured magnitude differences between PZ\,Tel\,B and its primary are listed for all observing epochs.}
\begin{tabular}{c|c|l|c}
\hline\hline
optics & epoch & filter & $\Delta m$ [$mag$]\\
\hline
S13 & 2010/05/05 & J                       & \,$5.54\pm0.20$\\
S13 & 2010/05/06 & $\rm J+ND_{Short}$      & \,$5.76\pm0.13$\\
S13 & 2010/05/07 & $\rm J+ND_{Short}$      & \,$5.69\pm0.16$\\
\hline
S13 & 2010/05/05 & $\rm H+ND_{Short}$      & \,$5.53\pm0.12$\\
S13 & 2010/05/07 & $\rm H+ND_{Short}$      & \,$5.49\pm0.14$\\
\hline
S27 & 2007/06/13 & $\rm K_{s}+ND_{Short}$  & \,$5.37\pm0.15$\\
S13 & 2009/09/28 & $\rm K_{s}$             & $>5.0\pm0.1$\\
S13 & 2010/05/05 & $\rm K_{s}+ND_{Short}$  & \,$5.33\pm0.09$\\
S13 & 2010/05/07 & $\rm K_{s}+ND_{Short}$  & \,$5.35\pm0.11$\\
\hline
\end{tabular}
\label{table_photo}
\end{table}

As PZ\,Tel\,A is saturated in our deep NACO images, taken in September 2009, only a lower limit for the magnitude-difference in the $\rm K_{s}$-band could be derived ($\Delta K_{s} > 5.0\pm0.1$\,mag). The determined limit agrees with the photometry of PZ\,Tel\,B obtained in all other observing epochs, in which NACO's neutral density filter $\rm ND_{Short}$ was used in the case that PZ\,Tel\,A would have saturated the NACO detector. The photometric measurements from the individual observing epochs are all consistent with each other within their uncertainties.

The apparent magnitudes of PZ\,Tel\,B and its primary can be derived with the obtained magnitude-differences, as well as the accurate photometry of the PZ\,Tel system ($J = 6.856 \pm 0.021$\,mag, $H = 6.486 \pm 0.049$\,mag, $K_{s} = 6.366 \pm 0.024$\,mag), which is listed in the 2MASS point source catalogue \citep{skrutskie2006}.

Finally, the precise Hipparcos parallax of PZ\,Tel yields a distance modulus $E=3.480\pm0.127$\,mag, which is used to derive the absolute magnitudes of PZ\,Tel\,B ($M_{J}=9.05\pm0.12$\,mag, $M_{H}=8.52\pm0.13$\,mag, and $M_{K_{s}}=8.24\pm0.10$\,mag), assuming that the NACO and 2MASS JH$K_{s}$ color systems
are identical.

\section{Limits on additional companions}

\begin{figure} [htb]
\resizebox{\hsize}{!}{\includegraphics{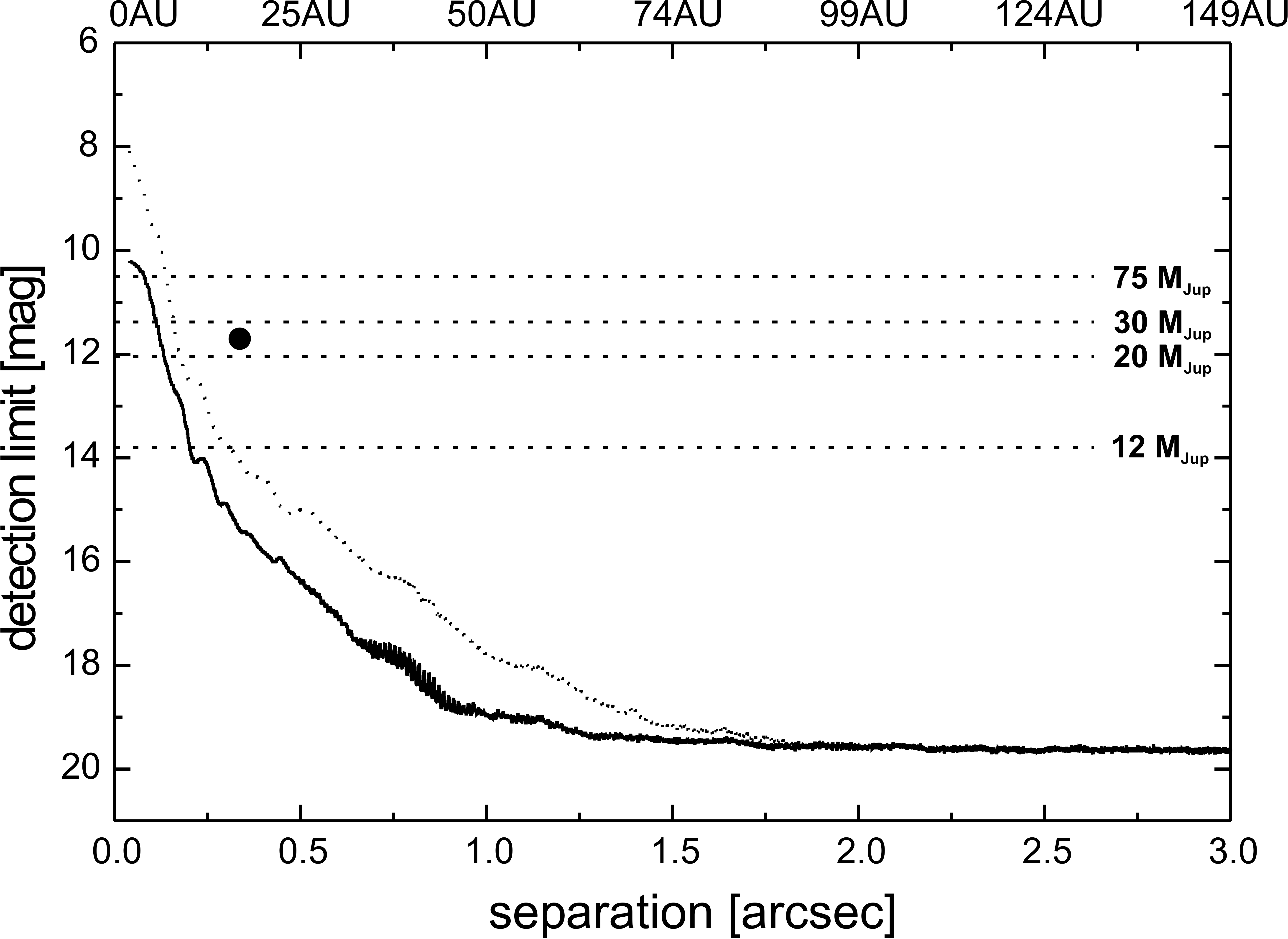}}\caption{The average detection limit of our deep NACO image taken in the $\rm K_{s}$-band in September 2009 plotted for a range of angular (bottom) and projected separation (top). The detection limit achieved in this image is shown as dotted line, as well as the detection limit after PSF subtraction (solid line). Due to saturation, objects within a separation of 3 NACO pixels (or 40\,mas of angular separation) cannot be detected around the bright primary. The position of PZ\,Tel\,B is indicated as black circle. Dashed horizontal lines show the expected brightness of substellar objects, whose masses are listed at the right side of the plot.}\label{limit}
\end{figure}

We obtained deep NACO observations of PZ\,Tel\,A and its companion in September 2009 in the $\rm K_{s}$-band. The achieved detection limit of our NACO image with and without PSF subtraction is shown in Fig.\,\ref{limit} together with the expected magnitudes of substellar objects with different masses at an assumed age of 10\,Myr, derived with the \cite{baraffe2003} evolutionary models and the well known distance of the PZ\,Tel system.

After PSF subtraction, all brown dwarf companions (mass$>12\,M_{Jup}$) are detectable in our NACO image beyond 0.23\,arcsec (or $\sim$\,11\,AU of projected separation) around PZ\,Tel\,A up to about 7.7\,arcsec ($\sim$\,380\,AU) at the outer edge of the field of view, fully covered by NACO's S13 optics (jitter-width taken into account). In the background noise limited region, beyond a separation of about 1.75\,arcsec ($\sim$\,87\,AU), a sensitivity of $K_{s} = 19.6$\,mag is reached in average, which allows the detection of planetary mass objects down to a mass of about $2\,M_{Jup}$ around PZ\,Tel\,A.

Beside PZ\,Tel\,B, one further very faint ($\Delta K_{s}>10.2$\,mag) companion-candidate is detected in our deep NACO image at $sep = 3.818\pm0.006$\,arcsec, and $PA=166.21\pm0.11^{\circ}$. This candidate was already detected by \cite{chauvin2010}, who proposed that it is most probably a background source (Epoch: 13/06/2007, $sep = 4.008\pm0.008$\,arcsec, $PA=166.3\pm0.2^{\circ}$, $\Delta K_{s}=10.7\pm0.1$\,mag). In the case that this object is a non-moving background object, we expect to find it at $sep=3.824\pm0.009$\,arcsec, and $PA=165.80\pm0.21^{\circ}$ in September 2009. Our NACO astrometry of the candidate agrees well with the predicted values, hence this source is a slowly moving object, clearly not related with the PZ\,Tel system. Hence, on basis of the achieved detection limit, we can conclude that there is no further companion of PZ\,Tel\,A within a separation of 7.7\,arcsec ($\sim$\,380\,AU) around the star. However, we cannot exclude additional very close companions. \citet{malaroda2000} list multi-epoch radial-velocity data of PZ\,Tel with a scatter of 4.4\,km/s, which could possibly be induced by such a close companion or by the spots on the stellar surface \citep{barnes2000}.

\section{Orbital motion}

Within all observing epochs, from June 2007 to May 2010, the separation and position angle of the PZ\,Tel\,B change both linearly over time with slopes of $\dot{sep} = 34.7\pm1.0$\,mas/yr and $\dot{PA} = -0.41\pm0.08$\,$^{\circ}$/yr. As PZ\,Tel\,B is located at an angular separation of $\sim\,$0.3\,arcsec in average, which corresponds to a projected separation of about 15\,AU at the distance of PZ\,Tel, the expected escape velocity is $\sim$\,46\,mas/yr, i.e. 1.3 times larger than the measured motion of the companion relative to its primary. Hence, the high relative motion of PZ\,Tel\,B (detected on a significance level $> 37\,\sigma$) is consistent with orbital motion of the companion around PZ\,Tel\,A. Between June 2007 and September 2009 the separation and the position angle of PZ\,Tel\,B change with slopes of $\dot{PA} = -0.5\pm0.3$\,$^{\circ}$/yr and $\dot{sep} = 35.8\pm1.3$\,mas/yr (see also section 2), while we measure slopes of $\dot{sep} = 30.9\pm2.3$\,mas/yr and $\dot{PA} = -0.2\pm0.4$\,$^{\circ}$/yr between September 2009 and May 2010, i.e. smaller absolute values of both velocities in the second time interval. The relative motion of the companion slightly slows down (detected on a $\sim2\,\sigma$ significance level), while its separation to its primary increases, consistent with orbital motion. Due to the slow variation in $PA$ compared with that of $sep$ the system is likely to be seen nearly edge-on.

With the obtained NACO detection limit we conclude that PZ\,Tel\,B can be imaged with NACO (even without PSF subtraction) down to a separation of 0.16\,arcsec around its much brighter primary. By tracing back the trajectory of the companion (derived from its relative motion) we expect that PZ\,Tel\,B should have been detectable with NACO only since September 2004. Indeed, PZ\,Tel\,A was observed with NACO in July 2003, as reported by \cite{masciadri2005} who also took deep $\rm K_{s}$-band images of the star, but could not detect its companion.

\section{Conclusions}

According to the evolutionary models for low-mass objects from \cite{chabrier2000}, \cite{baraffe2002} \& (2003) \nocite{baraffe2003} the absolute J, H, and $\rm K_{s}$-band magnitudes of PZ\,Tel\,B are consistent with a brown dwarf with $28^{+12}_{~-4}$\,$M_{Jup}$, at an age of $12^{+8}_{-4}$\,Myr, for the $\beta$\,Pic moving group. Even in the case that the PZ\,Tel system would be older than 20\,Myr the mass of PZ\,Tel\,B would still be within the substellar mass-regime, e.g. 69\,$M_{Jup}$ at 40\,Myr. The model predicted color $J-K_{s}$ for such a brown dwarf companion is $J-K_{s}=0.78\pm0.03$\,mag \citep[CIT$-$2MASS color transformation applied,][]{carpenter2001}, which agrees well with the color of PZ\,Tel\,B, derived from our NACO photometry $J-K_{s}=0.80\pm0.12$\,mag. According to the used models the effective temperature of PZ\,Tel\,B should range between 2500 and 2700\,K, which corresponds to a spectral type M6-8 \citep{golimowski2004}.

With only about 15\,AU of projected separation PZ\,Tel\,B is the second closest substellar companion imaged directly around a young star after $\beta$\,Pic\,b \citep{lagrange2010}, i.e. much closer than e.g. the substellar companions of GQ\,Lup \citep{neuhaeuser2005}, CT\,Cha \citep{schmidt2008}, or HR\,7329 \citep{lowrance2000, guenther2001}. With $\beta$\,Pic \citep{smith1984} and HR\,7329 \citep{smith2009}, PZ\,Tel is the youngest star with both a substellar companion and a debris disk, which are all members of the $\beta$\,Pic moving group, which shows a high rate of stars with both debris disks and substellar companions. PZ\,Tel\,B is the first substellar companion imaged directly, where orbital motion is detected (as deceleration).

\begin{acknowledgements}

We would like to thank the technical staff and all support astronomers at ESO Paranal Observatory. We made use of the 2MASS public data releases, as well as the SIMBAD and VIZIER databases, operated at the Observatoire Strasbourg. NV acknowledges the support by projects DIPUV 07/2007 and Gemini-CONICYT 32090027. RN \& TOBS would like to thank DFG for support in program NE515/30-1.

\end{acknowledgements}

\bibliography{ref}
\bibliographystyle{aa}

%\begin{thebibliography}{}
%\end{thebibliography}{}

\end{document}